\documentclass[10pt]{iopart}
\usepackage{epsfig}
\begin{document}

\title{Correlated hip motions during quiet standing}

\author{R.K. Koleva\dag, A. Widom\dag\footnote[3]
{To whom correspondence should be addressed.}, D. Garelick\dag , 
Meredith Harris\dag\dag , Claire Gordy\ddag\ and Mia Jackson\ddag\ }

\address{\dag\ Physics Department, Northeastern University, Boston MA 
02115, USA}

\address{\dag\dag\ Department of Physical Therapy, Northeastern University, 
Boston MA 02115, USA}

\address{\ddag\ Research Science Institute, Massachusetts Institute of 
Technology, Cambridge MA 02139, USA}

\begin{abstract}
Kinematic measurements of two simultaneous coordinates from postural 
sway during quiet standing were performed employing multiple ultrasonic 
transducers. The use of accurate acoustic devices was required for the 
detection of the small random noise displacements. The trajectory in the 
anteroposterior - mediolateral plane of human chest was measured and 
compared with the trajectory in anteroposterior direction from the upper 
and lower body. The latter was statistically analyzed and appeared 
to be strongly anti-correlated. The anti-correlations represent 
strong evidence for the dominance of hip strategy during an unperturbed 
one minute stance. That the hip strategy, normally observed for large 
amplitude motions, also appears in the small amplitude of a quite stance, 
indicates the utility of such noise measurements for exploring the 
biomechanics of human balance. 
  
\end{abstract}


\section{Introduction}

Physical investigations into the motor control of body motions have 
traditionally been of three types (Shumway-Cook and Wallacot 1995): 
(i) direct muscle tension measurements, (ii) kinetic analysis 
of forces, and (iii) kinematic analysis of body displacements. 
The experimental techniques used to study body motions reflect 
the requirements of the above types of investigation. For example, 
electromyography has been the technique most often used to monitor 
the activity of muscles (Granley 1984, Winter 1990, Perry \etal 1981). 
Electrodes are placed on the skin above the particular muscles of interest. 
Force transducers and force plates measure the ground body reaction forces 
to body movements (Prietro \etal 1993, Firsov \etal 1993, Winter 1990, 
Goldie \etal 1989).

In the present work, purely kinematic studies of body motions will be 
of interest. Previous studies have employed electrical potentiometers 
to measure joint angles in those cases for which angular increments induce 
voltage differences (Barlett \etal 1986, Campbell \etal 1989). Accelerometers 
have been constructed from force transducers which also (ultimately) 
depend on kinematic induced voltages (Thomas and Whitney 1959). Finally, 
video, cinematography and optoelectric systems have been used to form images 
of body motions. The optoelectric systems require that infrared sources 
and/or reflectors be worn on each anatomic landmark to be imaged 
(Whittle 1991, Winter 1990). 

Kinematic experimental methods have probed a well known set of 
strategies that are used by humans to maintain balance (Nasher 1985). 
The strategies involve the ankles, the hips and stepping. 
These three forms of motion are exhibited in various degrees 
depending on the nature of the motor task.

The purpose of this work is to introduce a new ultrasonic sound wave 
assessment (SWA) device which aids the analysis of body movement 
kinematics. The SWA device measures the coordinates 
\begin{math} X_1,X_2,... \end{math} of anatomic landmarks as a function 
of time. This device has been employed to investigate balance strategies 
during quiet standing. Surprising but unambiguous experimental results  
show that strongly correlated hip motions play a central role even in the 
postural sway of an {\em unperturbed stance}. This experimental result 
was made possible due to the accuracy of the SWA device. 

In Sec.2, the SWA device will be described in detail. In Sec.3, two 
coordinate measurements are discussed. The measurements involve the 
random noise motions of quietly standing subjects. Two coordinates 
were selected so as to emphasize the hip strategy. The observed restricted 
motions in the two coordinate phase space indicated the dominance of the 
hip strategy in the quiet postural stance. In Sec.4 we calculate the 
correlation functions for the two coordinate phase space and deduce that 
the hip motions are quite strongly correlated in {\em all of the subjects} 
measured. In the concluding Sec.5, previous notions of quiet standing 
balance strategies will be discussed in the light of presently reported 
data. 

\section{Sound wave assessment device}

The SWA device consists of an even number of small ultrasonic 
transducers. The measurement of \begin{math} \cal{N}  \end{math} 
coordinates requires \begin{math} 2\cal{N}  \end{math} transducers. 
Each of the \begin{math} \cal{N} \end{math}  {\em pairs} of transducers 
send ultrasonic pulses to one another at a rate of 
\begin{math} 1800 \end{math} pulses per minute. One transducer of each 
pair is positioned on a stable laboratory stand. The other transducer of 
the pair is attached to a quietly standing subject. Each pulse sent 
by one transducer in a given pair is later detected by the other transducer 
in the same given pair. The distance between the two transducers of a 
given pair  can then be obtained from the time required for the pulses to 
travel from the sender to the receiver. Thus, each pair of transducers 
measures one coordinate function \begin{math} X(t) \end{math}, i.e. the 
distance between the pair of transducers at time 
\begin{math} t \end{math}.      

In the photography industry, a single ultrasonic transducer system 
has been used to measure the distance between the camera lens and 
the object to be photographed. Twice the distance is computed from 
the measured pulse emission time and the measured pulse detection 
time at which the 
transducer detects its own echo. Our use of {\em transducer pairs} 
improves the accuracy of the coordinate measurement system by having 
less air absorption as well as less spatial dispersion (fanning out) 
of the ultrasound signals.

Each ``pulse''  (produced at the rate of \begin{math} 1800 \end{math} 
per minute) in reality consists of \begin{math} 16 \end{math} 
very closely spaced pulses separated by time intervals of 
\begin{math} \delta t\approx 2\times 10^{-5}\sec \end{math}. If 
a transducer pair is well aligned, then first pulse of the sixteen 
sub-pulses will be detected. If the transducer pair alignment is somewhat 
skewed, then the second sub-pulse will be detected. The coordinate 
displacements are measured to within \begin{math} 0.02\ cm \end{math} 
in a pulse bandwidth of \begin{math}\sim 12\ kHz \end{math}. 
This yields a displacement noise error of 
\begin{math}\delta X\sim 2\ (\mu m/\sqrt{Hz})\end{math}. 

Thus, we can record the fine displacements of a subject's body coordinates   
\begin{math} X_1(t),X_2(t), ... \end{math} as a function of time. Twelve  
healthy subjects in the age range between \begin{math} 15 \end{math} 
and \begin{math} 65 \end{math} years of age participated in this study.
 
\section{Measured paths in two coordinate planes}

Shown in Fig.1 is the random motion in the \begin{math}(X_A,X_M)\end{math} 
plane of a subject's upper body motion during quiet standing. The coordinate 
\begin{math} X_A \end{math} measures the subject's upper body anteroposterior 
displacement, i.e. the forward and backward  
motions of the transducer attached to the subject's back. The coordinate  
\begin{math} X_M \end{math} measures the upper body mediolateral 
displacement, i.e. the side to side motions of the transducer attached 
to the subject's shoulder. The random motions were measured over a 
time period of one minute. The quantities  
\begin{math} \Delta X_A=X_A-\bar{X}_A \end{math} and 
\begin{math} \Delta X_M=X_M-\bar{X}_M \end{math} denote 
the deviations from the mean taken over the observation time.
One notes the meandering nature of the noise in the 
\begin{math}(X_A,X_M) \end{math} plane.
\begin{figure}
\begin{center}
\epsfig{file=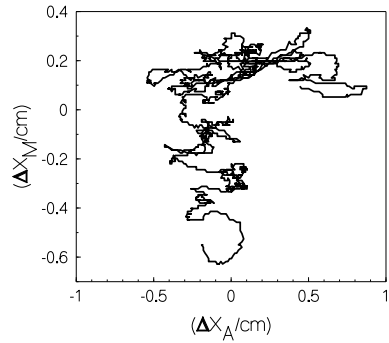,width=6cm}
\end{center}
\caption{A measured quiet standing upper body random meandering path 
in the plane of the  anteroposterior coordinate 
\begin{math}X_A \end{math} and mediolateral coordinate 
\begin{math}X_M \end{math} is plotted. The observation time is one 
minute. }
\end{figure}

\begin{figure}
\begin{center}
\epsfig{file=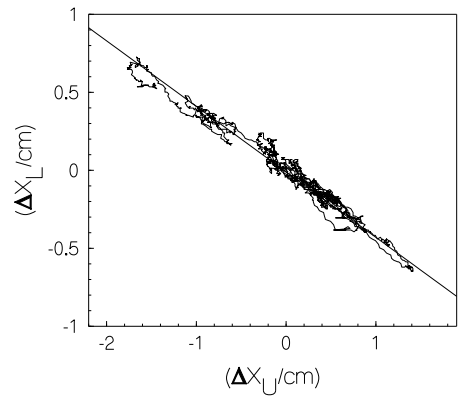,width=7cm}
\end{center}
\caption{A measured quiet standing highly correlated 
path in the plane of the upper body coordinate 
\begin{math}X_U \end{math} and lower body coordinate 
\begin{math} X_L \end{math} is plotted. The observation time 
is one minute. That the random path is almost {\em collinear} 
in the \begin{math}(X_U,X_L) \end{math} plane is evidence 
of the highly collective nature of the dominating hip strategy.}
\end{figure}

By contrast, in Fig.2, we consider the one minute measurement 
(on the same subject) in the \begin{math}(X_U,X_L)\end{math} plane 
with {\em both displacements} in the anteroposterior direction. 
The coordinate \begin{math}X_U=X_A \end{math} was again measured 
from a transducer attached to the subject's back (upper body) while 
the coordinate  \begin{math} X_L \end{math} was measured from a 
transducer attached to the back of the subjects thigh slightly below the 
hip (lower body). The deviations from the mean are 
\begin{math}\Delta X_U=X_U-\bar{X}_U \end{math} 
and \begin{math}\Delta X_L= X_L-\bar{X}_L\end{math}. One notes 
the very constrained nature of the noise in the 
\begin{math}(X_U,X_L) \end{math} plane. The coordinates 
\begin{math}X_U \end{math} are highly correlated. The positions in 
the plane are nearly {\em collinear}, as shown in Fig.2 along with 
a ``best fit'' line. 

\begin{figure}
\begin{center}
\epsfig{file=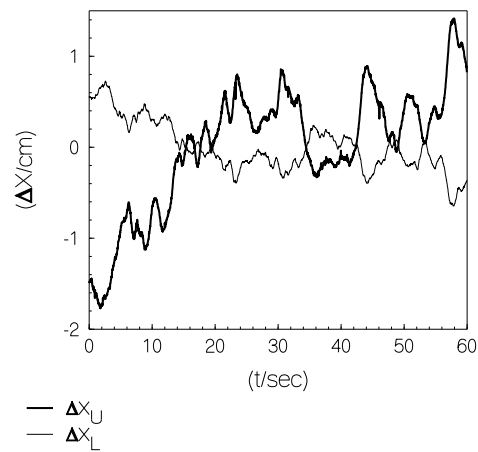,width=7cm}
\end{center}
\caption{Shown are plots of the two coordinates 
\begin{math}X_U(t) \end{math}
and 
\begin{math}X_L(t)\end{math}
as functions of time. This data was used to determine the path in 
Fig.2. One notes the strong anti-correlation between the coordinates. 
When \begin{math}X_U(t) \end{math} is increasing, 
\begin{math}X_L(t) \end{math} is decreasing in proportionate amounts, and 
vice versa.}
\end{figure}

In Fig.3, we plot both coordinates \begin{math}X_U(t) \end{math} and 
\begin{math}X_L(t)\end{math} as a function of time for 
an observation time of one minute. If one views only one of the 
coordinates, then the motion appears to be random. However when 
one views both coordinates, the changes in 
\begin{math}X_U(t) \end{math} and \begin{math}X_L(t)\end{math} 
are seen to be strongly anti-correlated. When one of the coordinates 
increases, the other coordinate decreases in proportionate amounts. 
Thus, when the upper body moves forward, the lower body moves backward 
and vice versa. This describes precisely the hip balance strategy. 

\section{Statistical analysis}

A quantitative statistical formulation of the anti-correlated 
kinematics of the hip strategy may be formulated as follows (Martin 1967): 
(i) For a given subject, the time averaged coordinate correlation matrix 
elements may be defined as 
\begin{eqnarray}
{\cal G}_{UU}={1\over \tau}\int_0^\tau 
\Delta X_U (t)\Delta X_U (t)dt, \\
{\cal G}_{LL}={1\over \tau}\int_0^\tau 
\Delta X_L (t)\Delta X_L (t)dt, \\
{\cal G}_{UL}={\cal G}_{LU}=
{1\over \tau}\int_0^\tau 
\Delta X_U (t)\Delta X_L (t)dt ,
\end{eqnarray} 
where \begin{math}\tau =60\ sec\end{math} is the observation time. 
(ii) The normalized cross correlation is defined 
\begin{equation}
{\cal C}_{UL}={{\cal G}_{UL}\over 
\left({\cal G}_{LL}{\cal G}_{UU}\right)^{1/2}}\ , 
\end{equation}
which obeys the inequalities 
\begin{math}-1\le {\cal C}_{UL}\le +1 \end{math}. The extreme 
of the inequality \begin{math} +1\end{math} indicates perfect 
correlation,  \begin{math}-1\end{math} indicates perfect anti-correlation 
and \begin{math}0\end{math} indicates lack of any correlation. 

{\em All of the measured subjects} exhibited a cross correlation 
function in the range 
\begin{math}0.89\le (-{\cal C}_{UL})\le 0.99 \end{math} 
indicating an almost perfect anti-correlation. For the totality of 
the experimental data, one may define the ensemble average of 
the correlation matrix elements for \begin{math} N=12 \end{math} 
subjects 
\begin{equation}
\bar{G}_{ab}={1\over N}\sum_{j=1}^{N}{\cal G}^{(j)}_{ab},
\ \ {\rm where}\ \  a=U,L\ ,\  b=U,L\ ,
\end{equation}
and the ensemble total cross correlation 
\begin{equation}
\bar{C}_{UL}={\bar{G}_{UL}\over 
\left(\bar{G}_{LL}\bar{G}_{UU}\right)^{1/2}}\ \ .
\end{equation}
For the totality of the experimental data from all of the subjects 
we find \begin{math} \bar{C}_{UL}=-0.92 \end{math} almost completely 
anti-correlated.

The above statistical analysis verifies objectively and quantitatively 
the view expressed in Sec.3. The hip strategy consists of a motion wherein 
the upper body and lower body randomly oscillate but in proportionately 
opposite directions. 

\section{Discussion}

In the study of condensed matter statistical mechanics, it is well known 
that one may tease out of the thermal equilibrium noise data many important  
mechanical parameters (Martin 1967). These parameters normally enter into 
the large dynamical displacements that would be present were those systems 
to be externally driven. Similarly, the notion has appeared in the 
literature that it would be possible to learn much about 
non-equilibrium motor biomechanics by merely observing the small 
displacement noise noise during quite standing (Lauk \etal 1998). 
This view certainly turns out to be true for the hip strategy employed 
to keep ones balance, as shown from the data reported here.

Non-equilibrium hip motions are evident, for example, to those who 
have observed an infant in that time interval {\em after} knowing how 
to crawl but {\em before} learning how to walk (Shumway-Cook 1995). 
When {\em not} very quietly standing, the infant exhibits large hip 
motions while trying to stand. Since (during this critical period in 
life) the transition rate \begin{math}\Gamma \end{math} for a fall 
(Koleva \etal 1999) obeys \begin{math}\Gamma > 0.05\ Hz  \end{math}, 
the infant often falls when the upper body coordinate is positive, 
either by sitting (backward fall) or returning to a crawl (forward fall). 
The non-equilibrium hip motions are equally evident in 
adults trying to balance on a tight rope with their forward direction 
normal to the tight rope direction. Since normal (non-circus trained) 
adults have a tight rope transition rate for a fall which also 
obeys \begin{math}\Gamma > 0.05\ Hz  \end{math}, the 
large hip strategy motions are clearly seen to exist somewhat before the 
inevitable fall onto a safety net. 

What has been shown in this work, is that the hip strategy motions 
are clearly in operation during a quite stance, {\em even though the 
amplitude of hip motions are too small to observed by the unaided 
eye}. With the aid of the ultrasound transducers in the SWA device, 
the two coordinate measurements clearly show an ongoing hip strategy in 
each and every subject tested. By a judicious choice of coordinates,  
other strategies should also yield to a quiet standing noise analysis 
technique.

\section*{References}
\begin{harvard}
\item[]Barlett S A, Maki B E, Fernie G R and Holliday P J 
1986 {\it Med. Biol. Eng. Comput.}
{\bf 24} p~219-22
\item[]Campbell A J, Borrie M J and Srears G F 1989 
{\it J. Gerantol.} {\bf 44} p~M112-7
\item[]Collins J J and DeLuca C J 1994 {\it \PRL} {\bf 73} p~764-7
\item[]Firsov G I, Rosenblum M G and Landa P S 1993 
{\it AIP Conference Proceedings 285:
Noise in Physical Systems and 1/f Fluctuations (St. Louis)} 
(New York: AIP Press) p~717-20
\item[]Goldie P A, Bach T M and  Evans O M 1989 
{\it Arch. Phys. Med. Rehabil.} {\bf 70} p~510-17
\item[]Granley J K 1984 {\it Phys. Ther.} {\bf 64} p~1831-8
\item[]Koleva R K, Widom A, Garelick D and Harris M 
1999 {\it cond-mat/9907311} 
\item[]Lauk M, Chow C C, Pavlik A E and Collins J J 
1998 {\it \PRL} {\bf 80} p~413-6
\item[]Martin P C 1967 {\it Proc. Int. Conf. on Many-Body Physics
(Grenoble)} (New York: Gordon and Breach) p~39-171 
\item[]Nasher L M and G McCollum 1985 {\it Behav. Brain Sci.} 
{\bf 8} p~135-72
\item[]Perry J, Easterday C S and Antonelli D J 
1981 {\it Phys. Ther.} {\bf 61} p~7-15
\item[]Prieto T E, Myklebust J B and Myklebust B M 
1993 {\it IEEE Trans. Rehab. Eng.} {\bf 1}
p~26-34
\item[]Shumway-Cook A and Wallacot M 1995 {\it Motor Control: Theory 
and Pracical Applications} (Baltimore: Williams \& Wilkins) p~119
\item[]Thomas D P and Whitney R J 1959 {\it J. Anat.} {\bf 94}
p~524-39
\item[]Whittle M 1991 {\it Gait Analysis: An Introduction} 
(Butterworth \& Heinemann) p~131-73
\item[]Winter D A 1990 {\it  Biomechaning and motor control of 
human movement} (New York: John Wiley \& Sons)

\end{harvard}

\end{document}